# Robust 3D multi-polar acoustic metamaterials with broadband double negativity

Hao-Wen Dong[1, 2†], Sheng-Dong Zhao[3†], Yue-Sheng Wang[4*], Li Cheng[1*], Chuanzeng Zhang[5]

[1]*Department of Mechanical Engineering, The Hong Kong Polytechnic University, Hong Kong, PR China*
[2]*Department of Applied Mechanics, University of Science and Technology Beijing, Beijing 100083, PR China*
[3]*School of Mathematics and Statistics, Qingdao University, Qingdao 266071, PR China*
[4]*Department of Mechanics, School of Mechanical Engineering, Tianjin University, Tianjin 300350, PR China*
[5]*Department of Civil Engineering, University of Siegen, D-57068 Siegen, Germany*

[†] **These authors contributed equally to this work.**
[*] **Corresponding authors: li.cheng@polyu.edu.hk (L. Cheng); yswang@tju.edu.cn (Y. S. Wang)**

**Abstract**
Acoustic negative-index metamaterials show promise in achieving superlensing for diagnostic medical imaging. In spite of the recent progress made in this field, most metamaterials suffer from deficiencies such as low spatial symmetry, sophisticated labyrinth topologies and narrow-band features, which make them difficult to be utilized for symmetric subwavelength imaging applications. Here, we propose a category of robust multi-cavity metamaterials and reveal their common double-negative mechanism enabled by multi-polar (dipole, quadrupole and octupole) resonances in both two-dimensional (2D) and three-dimensional (3D) scenarios. In particular, we discover explicit relationships governing the double-negative frequency bounds from equivalent circuit analogy. Moreover, broadband single-source and double-source subwavelength imaging is realized and verified by 2D and 3D superlens. More importantly, the analogical 3D superlens can ensure the subwavelength imaging in all directions. The proposed multi-polar resonance-enabled robust metamaterials and design methodology open horizons for easier manipulation of subwavelength waves and realization of practical 3D metamaterial devices.

**Introduction**

During the past two decades, choreographed artificial metamaterials [1-3] have been explored to manipulate electromagnetic [1], acoustic [2, 4] and elastic [5-7] wave propagations. Acoustic metamaterials (AMMs) can lead to diverse subwavelength functionalities [8], such as perfect absorption [9-10], negative refraction [11-14], phase modulation [15-21], etc., showing promise for applications in sound reconstruction [22], imaging [13-14, 18, 21, 23-24], energy harvesting [25-26] and cloaking [27-28], etc. Since double negativity is a cornerstone in many applications such as the super-resolution medical ultrasonic imaging, double-negative AMMs have been drawing persistent attention and have been synthesized through various means such as coupled Helmholtz resonators [29], membranes [30], combined membranes and Helmholtz resonators [31], space-coiling units [11-12] and macroporous silicone rubber microbeads [31]. Anisotropic [14, 33] AMMs can also realize the subwavelength imaging. However, their highly unsymmetrical microstructures make them difficult to be integrated at large-scale for symmetric imaging applications.

For subwavelength airborne sound, the detuned Helmholtz resonators can be designed to capture visible double negativity [29]. With the use of two coupled membranes, double negativity can also be achieved with

---

tunable monopole and dipole resonances [30]. Alternatively, combining membranes and Helmholtz resonators has been shown to enable double negativity in the ultra-low frequency range [31]. Furthermore, the space-coiling structures can exhibit double negativity only in a relatively narrow frequency range [11]. As for the structural forms of double-negative solid-air microstructures, two bottle-necking problems hamper the practical 3D subwavelength imaging beyond the proof-of-concept designs. On one hand, most previously reported space-coiling metamaterials suffer from the low spatial symmetry and sophisticated labyrinth topologies, which limit the practical applicability of double-negative AMMs for subwavelength imaging. Besides, our previous work showed that highly symmetric space-coiling metamaterials are nearly impossible to acquire double negativity [34]. On the other hand, the lack of novel universal double-negative mechanism and topological feature hinders the 3D broadband subwavelength imaging.

In this article, we design, fabricate and test a novel class of multi-cavity AMMs featuring broadband double negativity. A 2D multi-cavity system is demonstrated to be analogous to the inductor-capacitor circuits (LC) with intrinsic hybridization of quadrupole and dipole resonances. To validate the 2D multi-polar AMMs, we numerically and experimentally demonstrate the single-source and double-source acoustic imaging and accomplish the broadband subwavelength properties by the designed 2D superlens. Moreover, we construct a simple, highly-symmetric and intuitionistic 3D AMM by exploiting the typical multi-cavity topological features. Then, we reveal its double-negative mechanism of intrinsic hybridization of octupole and quadrupole resonances. In particular, we discover explicit relationships governing the double-negative frequency bounds of multi-polar resonances in both 2D and 3D AMMs. The unique relationship cannot only explain the observed double-negative properties but also provide a simple and universal guidance for the subsequent metamaterial design. Finally, we design a 3D superlens to realize the broadband subwavelength single-source and double-source imaging numerically and experimentally. The most important feature about this 3D superlens is the ability of ensuring the subwavelength imaging in all directions. The proposed design concept and the revealed multi-polar resonance mechanism open new horizons for the design of 3D metamaterial devices.

**Results**

**Design of 2D multi-cavity metamaterials.** Using our previous inverse-design methodology [34], we design and fabricate a 2D microstructure comprising five big solid blocks, four narrow air channels and four air regions, as shown in Figs. 1(a) and 1(b). The retrieved normalized effective mass density [11, 34-35] $\rho_{\text{eff}}/\rho_0$ and the bulk modulus $K_{\text{eff}}/K_0$ in Figs. 1(c) and 1(d) with $\rho_0$=1.29 kg/m$^3$ and $K_0$=149124 Pa being the material parameters of the background air confirm the double negativity within [1682.7 Hz, 2419.5 Hz]. Notably, the broadband negative range of $\rho_{\text{eff}}$ is much larger than that of $K_{\text{eff}}$; and consequently, the latter dominates the resulting double-negative range. Importantly, compared with the space-coiling AMMs [11-12, 35], the multi-cavity AMMs in Fig. 1 essentially reduce the structural requirements for double-negative formation. Figures 1(c) and 1(d) also show that the double negativity can be effectively maintained even when the viscous losses [35] are considered. Despite the obvious shift of the resonant frequencies, the double-negative range can be significantly extended. Of course, with a narrower air channel, the double negativity can be further improved on the premise of an acceptable transmission.

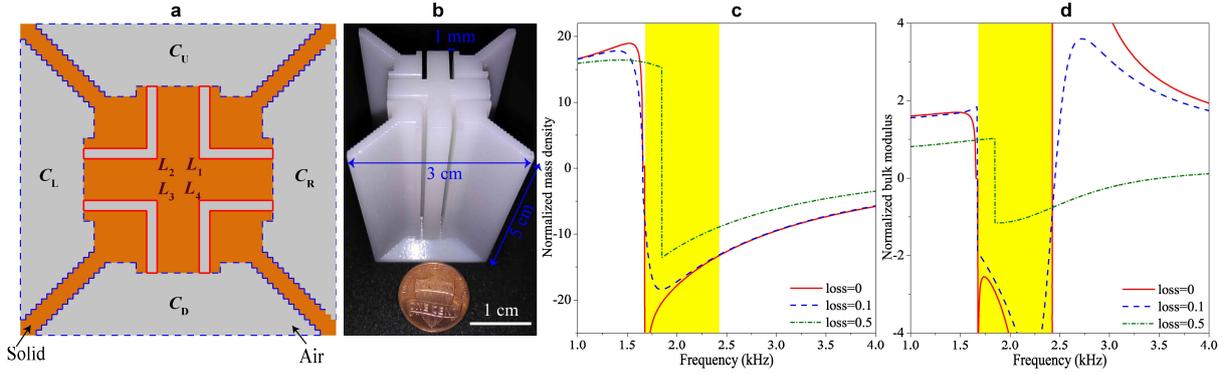

**FIG. 1. 2D multi-cavity AMMs. (a)** Design originating from our previous topology-optimized metamaterials [34]. **(b)** A fabricated microstructure. **(c)-(d)** Effective parameters with different viscous losses [35]. The shaded regions show the double-negative range irrespective of viscous loss (loss=0).

**Multi-polar mechanism of 2D multi-cavity metamaterials.** To understand the underlying physics of the observed double negativity, we scrutinize in Figs. 2(a)-2(c) three special eigenstates of AMMs in Fig. 1. Figure 2(a) clearly shows a typical quadrupole resonance behavior. The acceleration distribution is similar with the velocity distribution at **k**=0. The symmetric and diametrical distribution along *x*- and *y*- axes implies that the average velocity *v* over the microstructure is equal to zero, *i.e.* zero air flowing through the structure. Besides, the amount of air flowing into or out from the microstructure can be equivalent to $\partial V/\partial t$, *i.e.* $a \cdot v = \partial V/\partial t$, where $V$ and $a$ are the volume and lattice constant of the microstructure, respectively. Therefore $\partial V$ should be zero as well. In view of the basic definition $K_{\text{eff}} = -V \partial p/\partial V$, the resultant effective modulus should be infinite, where *p* denotes the acoustic pressure. Hence, the highly symmetric cavities can provide ideal structural building blocks for negative $K_{\text{eff}}$. From the opposite distributions in Figs. 2(b) and 2(c), it is recognized that the average particle acceleration ($\partial v/\partial t$) over the microstructure is negative, alongside a negative divergence of the pressure. Negative $\rho_{\text{eff}}$ can then be deduced from Newton's equation $-\nabla p = \rho_{\text{eff}} \partial v/\partial t$. Therefore, the eudipleural topological feature of the microstructure ensures the occurrence for a negative $\rho_{\text{eff}}$.

In view of the conspicuous resonant traits in Figs. 2(a) and 2(b), we can adopt the equivalent LC circuit [2] shown in Figs. 2(d) and 2(e) to analyze the resonant frequencies of the quadrupole and dipole resonances, respectively. To enable the analogy, the air domain in Fig. 1(a) is divided into four cavities and four air channels, marked by $C_L$, $C_R$, $C_D$, $C_U$, $L_1$, $L_2$, $L_3$ and $L_4$, respectively. For the eigenstate in Fig. 2(a), the acoustic pressure fields in the four cavities and channels form a typical quadrupole resonance. Based on the acousto-electrical analogy, the four air cavities and channels are equivalent to the four capacitors ($C_L=C_R=C_D=C_U=C_{2\text{eff}}$) and four inductors ($L_1=L_2=L_3=L_4=L_{2\text{eff}}$) in Fig. 2(d). For the dipole resonance in Figs. 2(b) and 2(c), the acoustic pressure is mainly accentuated in the left and right air cavities and four channels. So the equivalent circuit can be simplified as two capacitors and four inductors as illustrated in Fig. 2(e). As a result, the resonant frequencies of the quadrupole ($f_{2Q}$) and dipole ($f_{2D}$) resonances can be estimated by

$$f_{2Q} = \frac{1}{2\pi \sqrt{L_{2\text{eff}} C_{2\text{eff}} /4}}, \tag{1}$$

and

$$f_{2D} = \frac{1}{2\pi \sqrt{L_{2\text{eff}} C_{2\text{eff}} /2}}, \tag{2}$$

where $C_{2\text{eff}} = V_C a^2 / K_0$ with $V_C$ being the volume fraction of one cavity; $L_{2\text{eff}}$ is calculated by $L_{2\text{eff}} = \rho_0 l_L / (w_L)$ in which $l_L$ and $w_L$ denote the effective length and cross-sectional width of one channel normalized to the lattice constant, respectively. The predicted relative double-negative bandwidth is 34.3% which is very close to 35.9% extracted from Figs. 1(c) and 1(d).

Intriguingly, the two resonant frequencies are directly linked to each other through a constant of $\sqrt{2}$. This specific property is attributed to the typical feature of the multi-cavity AMMs with multi-polar resonances. We can then surmise that all symmetric multi-cavity metamaterials should have a broadband double negativity within the ascertainable frequency bounds. Figures 2(f) and 2(g) depict the variations of the quadrupole resonance frequency with three predominant parameters, $V_C$, $w_L$ and $l_L$. Clearly, narrow and long air channels combined with large cavities are beneficial to the low-frequency double negativity. When $l_L$ keeps in constant, $w_L$ can cause a greater impact on $f_{2Q}$ than $V_C$, see Fig. 2(f). However, both $V_C$ and $l_L$ can have conspicuous effect on $f_{2Q}$ when $w_L$ is given, see Fig. 2(g). Consequently, one can easily adjust these three decisive parameters to tune the double-negative property, thus avoiding the complex topology and multiple parameters involved in the space-coiling metamaterials [11-12].

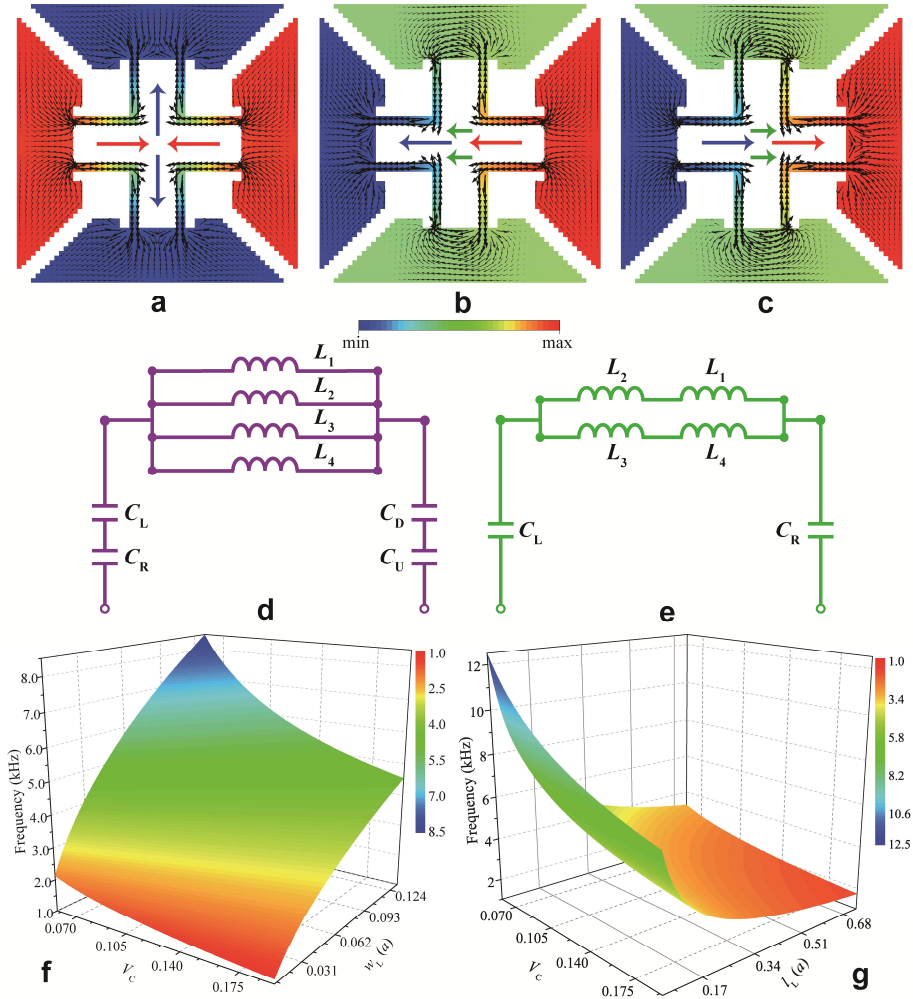

**FIG. 2. 2D double-negative mechanism. (a)** Eigenstate with acceleration distribution at 2419.5 Hz and **k**=(0, 0). **(b)** and **(c)** Eigenstates with acceleration (b) and velocity (c) distributions at 1682.7 Hz and **k**=($\pi/a$, 0). **(d)** and **(e)** Equivalent LC circuits for the eigenstates in (a) and (b). **(f)** and **(g)** Predicted quadrupole resonance frequencies by the circuit in (d) with three decisive parameters $V_C$, $w_L$ and $l_L$. Values of $l_L$ in (f) and $w_L$ in (g) are 0.43$a$ and 0.033$a$, respectively.

**Demonstration of broadband subwavelength imaging of 2D metamaterials.** To show the potential of the proposed multi-cavity topologies, we demonstrate the broadband single-source and double-source subwavelength imaging by the designed 2D AMM. Figures 3(a) and 3(b) depict the experimental setup with a metalens where the nearly closed air cavities distribute symmetrically around the solid cross blocks.

We first investigate the single-source imaging. Figure 3(c) shows the broadband feature of measured imaging. Notably, one can tactically adjust the geometrical sizes of the AMM to achieve a broader and lower frequency range. The measured acoustic field at 1950 Hz in Fig. 3(d) clearly shows the high-transmission (~35%) imaging. Undoubtedly, the imaging is ascribed to the double negativity. Moreover, most full widths at the half maximum (FWHM) of measured images as displayed in Fig. 3(e) can break the diffraction limit of $0.5\lambda_0$ ($\lambda_0$ refers to the acoustic wavelength in air) within the measured frequency range, thus manifesting the broadband subwavelength nature. Secondly, similar strategy is used to explore the double-source imaging. Two images are effectively separated in Fig. 3(f), thereby forming a subwavelength imaging resolution of $0.34\lambda_0$. To validate the stable imaging ability, we also check the imaging performance by varying the phase difference between the two sources. In fact, two images can be effectively distinguished over the large phase-shift range. The detailed evidentiary results can be found in the Supplemental Information. All measured imaging results also demonstrate that the viscous losses can hardly affect the double-negative properties, in agreement with the results in Figs. 1(c) and 1(d).

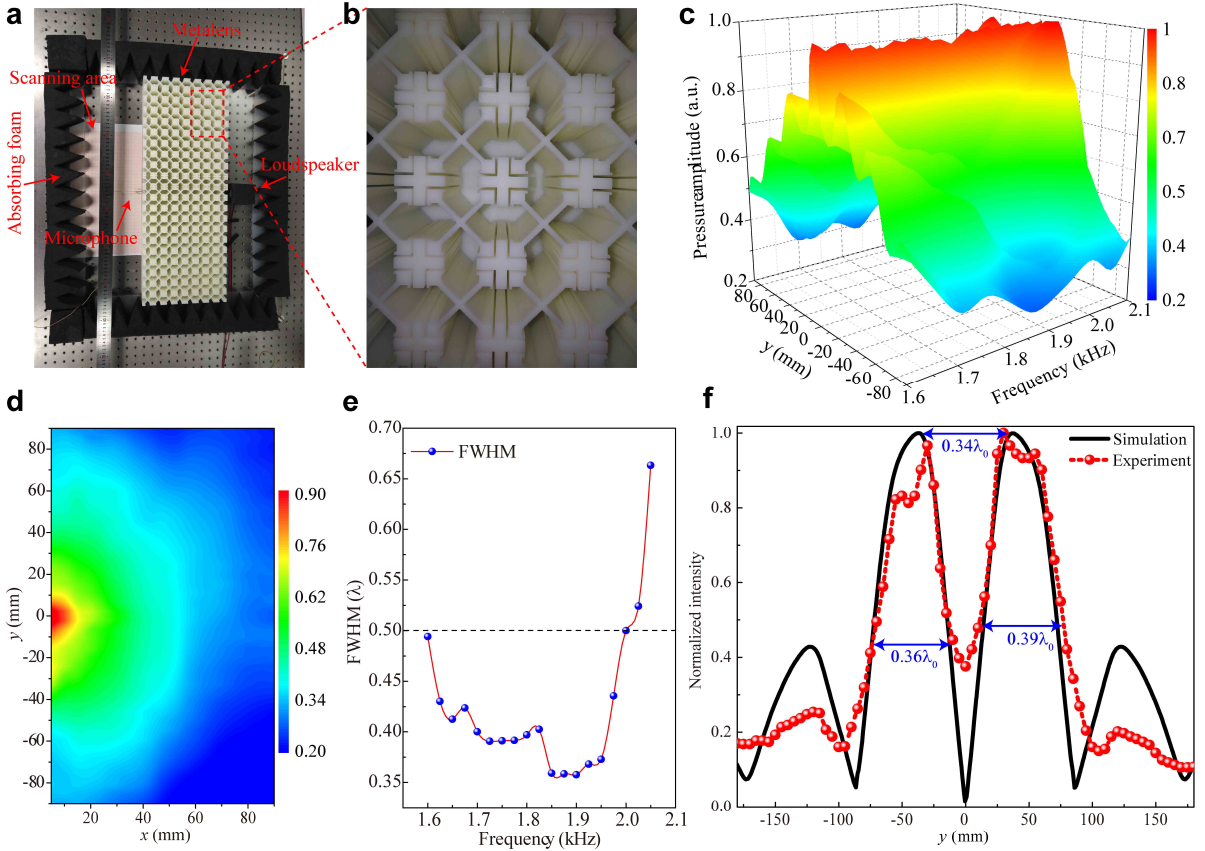

**FIG. 3. Subwavelength imaging using 2D AMMs.** (**a**) Experimental setup. (**b**) Local image of the superlens comprising 21×8 microstructures. (**c**) Measured pressure amplitude along the exiting surface of the metamaterial slab within [1600 Hz, 2100 Hz]. (**d**) Measured acoustic amplitude field inside a scanning domain of 9 cm×18 cm. (**e**) Measured single-source imaging resolutions within the double-negative range. (**f**) Simulated and measured sound intensity profiles of the imaging under two-point sources (6 cm apart) with a π phase shift at 1950 Hz.

**Analogical 3D multi-cavity metamaterials.** Since the 2D multi-cavity topologies with multi-polar resonances are the intrinsic topological features, as long as we capture these features, it should be straightforward to construct the analogical 3D double-negative AMMs with the quite simple, highly-symmetric and intuitionistic geometries. We can then demonstrate the broadband single-source and double-source subwavelength imaging as well. The analogical 3D AMM is composed of twelve symmetric air cavities, twenty-four air channels and one solid block, see Figs. 4(a)-4(c). Each facet of the 3D microstructure has a similar topological feature as the 2D model in Fig. 1. As demonstrated below, the proposed multi-polar resonance-enabled topologies really can ensure the double negativity. Therefore, the multi-polar mechanism and multi-cavity topology are robust for double-negative solid-air AMMs.

**Multi-polar mechanism of 3D multi-cavity metamaterials.** For explaining the double negativity, Figs. 4(d) and 4(e) show the octupole and quadrupole resonances, respectively. Due to the highly symmetrical distribution of sound pressure in Figs. 4(d), it is noted that the average velocity over the microstructure is equal to zero, leading to an infinite $K_{eff}$. However, in view of the symmetrical pressure distribution in Fig. 4(e), zero acceleration can be generated for the entire structure. More analyses about the dispersion relations and representative eigenstates are given in the Supplemental Information. Therefore, the resultant $\rho_{eff}$ should be infinite. Consequently, the hybridization of the quadrupole and octupole resonances endures a 3D double negativity.

Similarly, we can use an equivalent LC circuit to characterize the resonant behavior. For the octupole resonance, eight air cavities and twenty-four tunnels dominate the acoustic pressure filed and form the typical octupole resonance. So the cavities and tunnels are equivalent to the eight capacitors ($C_1=C_2\ldots=C_8=C_{3eff}$) and twenty-four inductors ($L_1=L_2\ldots=L_{24}=L_{3eff}$), respectively. For the quadrupole resonance, only four cavities and sixteen tunnels are heavily loaded by the acoustic pressure. So the equivalent circuit has only four capacitors and sixteen inductors. As a result, the resonant frequency of the 3D octupole ($f_{3O}$) and quadrupole ($f_{3Q}$) resonances can be predicted, respectively, by

$$f_{3O} = \frac{1}{2\pi\sqrt{L_{3eff}C_{3eff}/6}}, \tag{3}$$

and

$$f_{3Q} = \frac{1}{2\pi\sqrt{L_{3eff}C_{3eff}/4}}, \tag{4}$$

where $C_{3eff}$ and $L_{3eff}$ are homoplastically defined as $C_{3eff} = V_C a^3/K_0$ and $L_{3eff} = \rho_0 l_L/(\pi r_L^2)$, respectively, with $V_C$ being the volume fraction of one cavity; and $l_L$ and $r_L$ are the effective length and radius of one tunnel, respectively. The relative double-negative bandwidth is predicted to be 20.2%, close to 23.4% extracted from the dispersion relations. In particular, a fixed relationship of $\sqrt{1.5}$ between $f_{3O}$ and $f_{3Q}$ is observed. This emphasizes again the unique rule of the multi-polar resonance-enabled multi-cavity topologies even for 3D metamaterials. If other types of multi-cavity topologies are used, a fixed homologous relationship is also expected. According to Eqs. (3) and (4), three decisive parameters $V_C$, $r_L$ and $l_L$ dominate the double-negative range. Similarly, the larger cavities connected with the narrower and longer channels are favorable to double negativity at the lower frequencies. Compared with the 2D system, the predicted 3D frequency can be modulated to a much lower extent. More detailed results can be found in the Supplemental Information. Obviously, for both 2D and 3D systems, the proposed multi-polar resonance-enabled multi-cavity topology is suitable for constructing arbitrary broadband double-negative solid-air AMMs.

**Demonstration of broadband subwavelength imaging of 3D metamaterials.** The subwavelength single-source and double-source imaging can also be realized by the proposed 3D AMM, which is verified numerically and experimentally. Figure 4(f) clearly shows the ideal broadband high-transmission (~35%) imaging. In particular, the thickness of the three layers is already enough for the imaging. This idiosyncrasy greatly improves the effectiveness of the multi-cavity topologies. For the double-source imaging, the measured acoustic field in Fig. 4(g) confirms the imaging effect well. Meanwhile, the simulated and measured intensity profiles demonstrate the subwavelength resolution of $0.46\lambda_0$. The details are elucidated in the Supplemental Information. Moreover, whatever the phase shift is, two well separated foci can be obtained effectively as shown in Fig. 4(h). More details of simulated and measured single-source and double-source imaging using 3D multi-polar AMMs are presented and discussed in the Supplemental Information. Because of the high symmetry, the 3D AMM can safeguard the realization of imaging for acoustic waves perpendicularly incident on every facet of the sample. Certainly, the higher symmetric 3D AMM is easy to be constructed as long as the multi-polar resonance-enabled topological features are captured. Furthermore, it is expected to perform 3D topology optimization and generate the deeply practical resonance mechanism based on the multi-polar 3D AMMs in the future. These advantages will lay the foundation of deep-subwavelength ultrasound imaging.

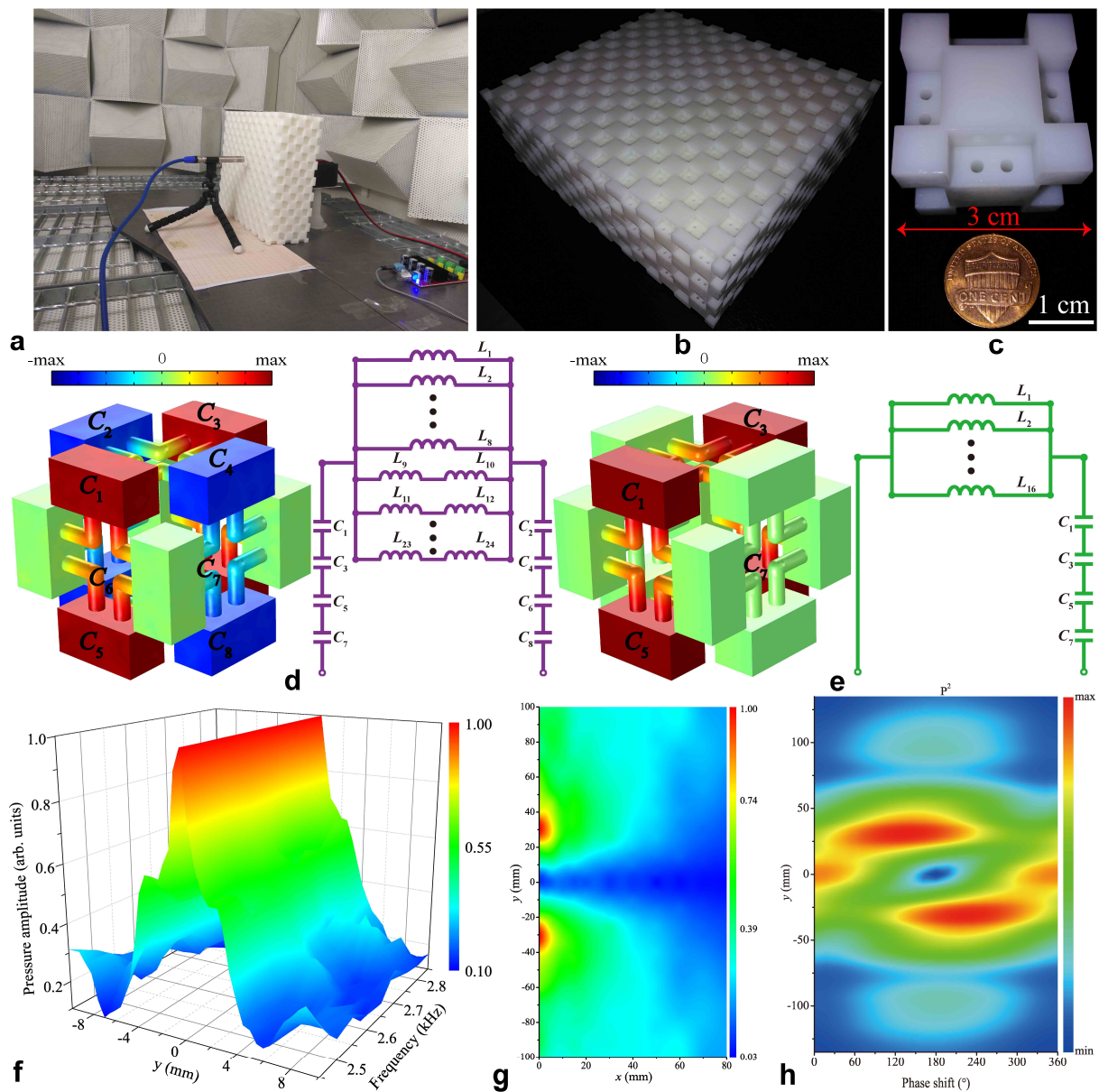

**FIG. 4. Analogical 3D multi-cavity AMMs and various evidences on broadband double negativity. (a)** 3D experimental setup. **(b)** Photograph of a 3D superlens containing 8×9×3 microstructures. **(c)** Fabricated 3D microstructure sample. **(d)** and **(e)** Eigenstates and their equivalent LC circuits with **k**=(0, 0, 0) at 3083.4 Hz (d) and with **k**=(0, $\pi/a$, 0) at 2436.5 Hz (e). **(f)** Measured acoustic pressure amplitude field along the exiting surface for a single source within [2450 Hz, 2850 Hz]. **(g)** Measured acoustic pressure amplitude field for two sources (6 cm apart) with a $\pi$ phase shift at 2650 Hz. **(h)** Simulated map of the intensity distribution in the focal plane with the phase shift between the two sources from $0°$ to $360°$.

## Discussions

In conclusion, we propose a kind of multi-polar resonance-enabled robust topologies to construct both 2D and 3D broadband double-negative AMMs. Benefiting from the high symmetries, the 2D AMM can support combination of quadrupole and dipole resonances. Similarly, the extension of the proposed topologies to 3D enables hybridization of octupole and quadrupole resonances. For either 2D or 3D AMMs, an explicit relationship governing the double-negative frequency bounds of multi-polar resonances is revealed. This discovery can be used as a universal criterion for double-negative multi-cavity metamaterial design. Finally, we demonstrate the broadband single-source and double-source subwavelength imaging of 2D and 3D superlens. The relatively simple, highly-symmetric, intuitionistic and robust multi-cavity topologies based on the unified multi-polar resonance mechanism are expected to boost the practical realization of the broadband superlens. The proposed design and methodology offer possibilities for airborne sound modulation using high practical 3D metamaterials [36].

## Methods

**Numerical simulations.** Simulations of the dispersion relations, effective constitutive parameters [11-12, 34-35] and eigenstates were performed by the commercial finite element software ABAQUS 6.14-1. Simulations of subwavelength acoustic imaging through 2D and 3D superlens were conducted by COMSOL Multiphysics 4.4.

**Experiments of 2D and 3D metamaterials.** In both 2D and 3D scenarios, we employed 3D printing to fabricate the microstrucutres and the metalens consisting of polylactice acid (PLA) with a mass density of 1250 kg/m$^3$ and bulk modulus of 3.5×10$^9$ Pa. To avoid acoustic reflections, the 2D sample was surrounded by the acoustic absorbing foams. However, the 3D acoustic experiments were conducted in a semi-anechoic room. A loudspeaker was taken as a single point source and located a certain distance from the input interface of the metalens, while the mounted microphone recorded the acoustic pressure over the entire scanning area. For the double-source experiments, two loudspeakers with a given phase delay and distance were used asdouble point sources. Signals obtained at each position in the scanning area were averaged over four measurements. The whole acoustic filed was obtained by using the Fourier transform after the scanning measurement.

## Acknowledgements

This work is supported by the Hong Kong Scholars Program (No. XJ2018041), National Natural Science Foundation of China (Grant Nos. 11802012 and 11532001), Project funded by China Postdoctoral Science Foundation (2017M620607), the Fundamental Research Funds for the Central Universities (Grant No. FRF-TP-17-070A1) and the Sino-German Joint Research Program (Grant No. 1355) and the German Research Foundation (DFG, Project No. ZH 15/27-1). We would like to thank Prof. Bilong Liu (Qingdao University of Technology, PR China) for his helpful participation.

## Authors Contributions

H. W. D conceived the origin idea. L. C and Y. S. W supervised the project. H. W. D and S. D. Z conducted

the physical analyses and simulations. S. D. Z performed the experiments. H. W. D wrote the manuscript. All authors discussed the results and commented to the manuscript.